\documentclass[12.0pt,letterpaper]{article}
\usepackage{comment}
\usepackage{graphicx}
\usepackage{bm}
\usepackage{amsthm} 
\newcommand{\vect}[1]{\vec{#1}} 
\usepackage{mwe} 
\usepackage{caption}
\usepackage{color}
\usepackage{lipsum}

\usepackage{graphicx}
\usepackage{amsmath,amssymb}
\usepackage[body={6in,8.5in}]{geometry}
\usepackage{color}
\usepackage{url}
\usepackage{authblk}
\usepackage{caption}
\usepackage{caption}
\usepackage{cleveref}
\usepackage{array}
\usepackage{chngcntr}

\newtheorem{theorem}{Theorem}[section]
\newtheorem{lemma}[theorem]{Lemma}

 \usepackage{mathptmx}

\title{Minor climatic fluctuations lead to species extinction in a conceptual ecosystem model}

\author[1,2]{Sergey A. Vakulenko}
\affil[1]{Institute for Problems in Mechanical Engineering, Russian Academy of Sciences, Bolshoy pr. V.O., 61, St. Petersburg 199178, Russia}
\affil[2]{Laboratory of Power Electronics and Automated Electric Drive, ITMO University, Kronverkskiy pr., 49, St.Petersburg 197101, Russia}

\author[3]{Ivan Sudakov \thanks{isudakov1@udayton.edu}}
\affil[3]{Department of Physics, University of Dayton, 300 College Park, Dayton, Ohio 45469, USA}

\author[4]{Luke Mander}
\affil[4]{School of Environment, Earth and Ecosystem Sciences, The Open University, Milton Keynes MK-7 6AA, UK}

\begin{document}
\maketitle







\begin{abstract}
The extinction of species is a core process that affects the diversity of life on Earth. One way of investigating the causes and consequences of extinctions is to build conceptual ecological models, and to use the dynamical outcomes of such models to provide quantitative formalization of changes to Earth's biosphere. In this paper we propose and study a conceptual resource model that describes a simple and easily understandable mechanism for resource competition, generalizes the well-known Huisman and Weissing model, and takes into account species self-regulation, extinctions, and time dependence of resources. We use analytical investigations and numerical simulations to study the dynamics of our model under chaotic and periodic climate oscillations, and show that the stochastic dynamics of our model exhibit strong dependence on initial parameters. We also demonstrate that extinctions in our model are inevitable if an ecosystem has the maximal possible biodiversity and uses the maximal amount of resources. Our conceptual modeling provides theoretical support for suggestions that non-linear processes were important during major extinction events in Earth history.



\end{abstract}

\maketitle

\section*{Lead Paragraph}
\begin{quotation}
{\bf The history of life on Earth is one of continually fluctuating diversity. In general terms the diversity of life, as measured by the number of species or higher taxa such as genera or families, represents the balance between the process of speciation (which adds species to the biosphere) and extinction (which removes species from the biosphere). Palaeobiological work has shown that the history of life is characterized by many extinction events that have at various times decimated the Earth's biota. Well-known examples of extinction events in Earth history include the event that destroyed the dinosaurs, and the process of extinction is of particular current scientific interest because it is thought that we may be approaching a mass extinction driven by anthropogenic activities. In this paper, we have developed a new conceptual ecosystem model that allows us to investigate how chaotic and period oscillations in Earth's climate affect biodiversity and extinction. This model is an extension of the well-known Huisman and Weissing model that has been used to study phytoplankton. Our model accounts for species self-regulation, extinctions, and time dependence of resources. The stochastic dynamics of our model are strongly dependent on initial parameters, and using analytical investigations and numerical simulations we show that non-linear processes are likely to be important aspects of extinction. In our model, extinctions are inevitable if an ecosystem has the maximal possible biodiversity and uses the maximal amount of resources. Our conceptual modeling provides a quantitative framework in which to investigate the dynamics of biospheric change.}
\end{quotation}

\section{Introduction}

The current state of the biosphere is a product of the evolutionary process that began with the origin of life around 3.5Ga.\cite{Wa80} Since this time, life has expanded from a single common ancestor to the diversity of biological forms that are present on the Earth today.\cite{Ben07} However, the diversification of life over this time interval has not been smooth or steady, and the fossil record indicates that there have been periods where the number of taxa has declined rapidly. Such intervals represent extinction events, and reviews of the history of life indicate that there have been 61 such events in Earth history.\cite{Wal96,Bam06} Of these, several stand out for their sheer magnitude.\cite{Ben95} These are mass extinctions, which are defined as "any substantial increase in the amount of extinction (i.e., lineage termination) suffered by more than one geographically wide-spread higher taxon during a relatively short interval of geologic time, resulting in an at least temporary decline in their standing diversity"(Sepkoski (1986), p. 278).\cite{Sep86}

 Palaeobiological studies indicate that extinction events are frequently associated with major environmental change. For example, several of Earth's largest extinction events occur during intervals of elevated volcanic activity, either due to the intrusion of large igneous bodies of rock as in the case of the Toarcian extinction event,\cite{El05} or to the opening of the Atlantic ocean in the case of the late Triassic extinction.\cite{Sch10} There are also examples of extinction events on much more recent timescales, such as the disappearance of the spruce tree species {\it Picea critchfieldii} during the last deglaciation in North America.\cite{Jack99, Man14} Such studies can provide empirical data on the sensitivity of the Earth's biota to environmental change, and can identify factors that can lead to the proliferation of species as well as the broad abiotic conditions under which species are lost from the Earth's biota.
	
A general trait that emerges from empirical palaeobiological studies of the biosphere is that extinctions reflect perturbations that stress ecosystems beyond their resilience.\cite{Sol02} Ecosystems represent functional entities that are produced by assembly processes, and if they are subject to perturbations that are greater in magnitude or duration than they can accommodate, then they are disrupted in some way.\cite{Mar68,Sol02} Conceptual ecological models do not represent every single complex biotic and abiotic interaction that leads to ecosystem assembly and disruption, but nevertheless, the dynamical outcomes of such models\cite{Plo93,Sol02} can provide a quantitative formalization for dynamical biospheric change, \cite{Sol02} and can serve as a counterpoint to empirical studies of biospheric evolution based on observational data.   

In this paper, we consider a new conceptual ecosystem model that not only gives a quantitative formalization but also allows us to investigate how biodiversity affects the mechanism of extinction. In this model, a number of species share resources, and oscillations in these resources (as might be induced by climatic change, for example), self-limitation effects, as well as extinctions are accounted for. Our conceptual model represents an extension of the Huisman and Weissing model,~\cite{HuWe99} which accounts for extinctions but only includes a single resource and does not include any climatic variation. The parameters of our model depend on the state of the environment via time dependent coefficients. This system is inspired by some phytoplankton models,~\cite{Hu61,Hu2005} and under certain assumptions can be derived from them, and if the resource turnover rate is large enough our model reduces to a Lotka-Volterra system.~\cite{KoVa}

Our paper is organized as follows: (1) we first state the standard model of species coexistence; (2) we then extend the standard model of species coexistence by introducing extinctions and climate, and assuming that the parameters depend on some environmental forcing that can oscillate (for example temperature); (3) we then consider the problem of extinction in our extended model in more detail; (4) finally, we use analytical investigations and numerical simulations to study the dynamics of our extended model under chaotic and periodic climate oscillations. Our principal results are that the stochastic dynamics of our model exhibit strong dependence on initial parameters. We show that extinctions are inevitable if an ecosystem has the maximal possible biodiversity and uses the maximal amount of resources, a conclusion which underscores the importance of studying the role of stability thresholds in mass extinctions.\cite{Rothman}

\section{The standard model of species coexistence} 

We consider the following standard model of population of biodiversity:~\cite{HuWe99}
\begin{equation}
     \frac{dx_i}{dt}=x_i (- r_i  + \phi_i(v) -  \sum_{j=1}^N \gamma_{ij} \; x_j),
    \label{HX2}
     \end{equation}
\begin{equation}
     \frac{dv_j}{dt}=D_j(S_j -v_j)   -  \sum_{k=1}^N c_{jk} \; x_k  \; \phi_k(v),
    \label{HV2}
     \end{equation}
where $v=(v_1, v_2, ..., v_M)$,   and
\begin{equation}
      \phi_j(v)= \min \{ \frac{a_j v_1}{K_{1j} + v_1}, ...,  \frac{a_j v_M}{K_{Mj} + v_M}  \} \, .
\label{MM2}
     \end{equation}
where $a_j$ and $K_{ij} >0$.
The terms $\gamma_{ii}x_i$ define
self-regulation of species populations that
restrict the species abundances, 
and $\gamma_{ij}x_j$ with $i \ne j$ define a possible competition  between species for resources. 
The coefficients $a_i$ are 
specific growth rates and the $K_i$ are self-saturation constants. 
The coefficients $c_{jk}$ determine how the 
species share the resource (nutrient supply). 


This model is widely used for primary producers like phytoplankton and it can also 
be used to describe competition among terrestrial plants.\cite{Til77}
When $\gamma_{ij}=0$ this system is equivalent to models used to study the plankton paradox, which describes the phenomenon where a limited range of resources supports an unexpectedly large number of different species.~\cite{HuWe99} 

Relation (\ref{MM2}) corresponds to the von Liebig minimum law, but we can consider even
more general $\phi_j$  satisfying  the conditions
\begin{equation}
      \phi_j(v) \in C^1, \quad    0 \le \phi_j(v) \le  C_+,  
\label{MM2a}
     \end{equation}
where $C_{+} >0$ is a positive  constant, and
\begin{equation}
      \phi_k(v) =0,   \quad  \forall k \quad v \in \partial {\bf R}^N_{>}     
\label{MM2b}
     \end{equation}
where $\partial {\bf R}^N_{>} $ denotes the boundary of the positive cone $ {\bf R}^N_{>} =\{v:  v_j \ge 0, \ \forall j\}$.
Note that condition (\ref{MM2b}) holds if $\phi_j$ are defined by (\ref{MM2}), so our conditions can be considered a a generalization of the von Liebig law, when the species abundance growth  stops if at least one resources vanishes. Thus, each resource is necessary for species survival.



We consider the Cauchy problem for the
system (\ref{HX2}),  (\ref{HV2}) on a time interval $[0, \tau]$, where initial data are given by
\begin{equation}
     x_i(0)=X_i >0, \quad v_k=V_k \in (0, S_k), 
    \label{Idata}
     \end{equation}
and $\tau$ is a positive number. 
We assume that $N >> 1$ (the case of a large ecosystem) and $X_i >0$ 
is distributed randomly according to a log-normal law with parameters 
$a, \sigma$. The corresponding distribution density is given by 


\begin{equation}
f(x)=\frac{1}{x\sigma\sqrt{2\pi}}\exp\left [ \frac{-{(\ln{x}-a)^2}}{2\sigma^2} \right ]
\label{law}
\end{equation}


Suppose we simplify the problem that 
 concurrence is absent and 
\begin{equation}
  \gamma_{ij}= \gamma_i \delta_{ij}, \quad \gamma_i >0.
\label{gam1}
\end{equation}
One can show, by standard estimates, that the Cauchy problem 
(\ref{HX2}), (\ref{HV2}) and (\ref{Idata}) is well posed and that
the corresponding $x_i(t)$ are defined for all $t >0$, bounded and take positive values. Therefore, we are dealing with a dynamical system. Moreover, we observe that this system is cooperative~\cite{Hir85} since
$\frac{\partial{F_i(\vect{x})}}{{\partial x_j}} >0$ for $j \ne i$.

\noindent
{\bf Assertion}. {\em Under condition (\ref{gam1})
the dynamical system defined by (\ref{HX2}), (\ref{HV2}) has a compact global attractor. In the case of a single resource ($M=1$) and sufficiently large 
turnover $D=D_1$ all trajectories of that system  are convergent, and there are no locally attracting stable limit cycles}. 

{\em Outline of the proof}.  We follow Kozlov et al. (2017) and Sudakov et al.(2017).\cite{Vak16, SVGK} The resource $R(\vect{x})$ is a uniformly bounded
function. This fact, in a standard way, implies uniform boundedness of $x_i(t)$ 
for large times $t$ and shows that  the
system (\ref{HX2}), (\ref{HV2}) defines a global semiflow, which
has an absorbing set. Thus, this semiflow 
is dissipative  and has a compact global attractor.  The claim on trajectories convergence 
follows from Theorem I in Kozlov et al. (2017).\cite{Vak16}

The problem can be simplified   for large turnovers
 ($D_k >> 1$).
Then one can show \cite{SVGK} that
systems (\ref{HX2}) and (\ref{HV2}) reduce to Lotka-Volterra systems of a special form.

\section{The model with extinctions}

We extend systems (\ref{HX2}) and (\ref{HV2}) to describe  two important effects.  The first effect is species extinctions, and in this section we focus on it.
 The second effect is a result of environmental influence on the dynamics of the modified systems (\ref{HX2}) and (\ref{HV2}). That effect will be considered in the next section.

 In reality abundances $x_i$ are discrete numbers, therefore, if the abundance becomes too small, the corresponding species must become extinct. To describe this effect mathematically,  we introduce a parameter $\delta >0$ and suppose that if the $i$-th species abundance $x_i(t)$ becomes less than  $\delta$, 
i.e., $x_i(t_0)=\delta$ and $\frac{dx_i(t_0)}{dt} < 0$ for some $i$ and $t_0 >0$, then the corresponding species
should be excluded from systems (\ref{HX2}) and (\ref{HV2}). We then set formally that $x_i(t) \equiv 0$ for all  $t > t_0$. For the case of a single resource this extended model is proposed and investigated in more detail in Kozlov et al. (2017).\cite{Vak16}

Note that after this modification the model stays mathematically well posed. \cite{Vak16}
Next, we introduce a function $N_{e}(t)$,  which is the number of surviving species at time $t$, i.e. the number of the indices $i$ such that
$x_i(t)  > \delta$. It is clear that  $N_{e}(t)$ is a piecewise constant non-increasing function.  Let $t_0 < t_1 < ...  < t_m < ...$ be the  points of discontinuity of this function.  
Within the intervals  $[t_k, t_{k+1}]$ the Cauchy problem for systems (\ref{HX2}) and (\ref{HV2}) is well posed, and therefore the Cauchy problem is well posed for the modified systems (\ref{HX2}) and (\ref{HV2}) with extinctions. There are two possible situations.  If $\lim_{t \to +\infty}
N_{e}(t)=N_{\infty}=0$, then all the species vanish.  If $N_{\infty} >0$, then on some infinite semiaxis $(t_m, +\infty)$ the modified system is equivalent to
 model (\ref{HX2}) and (\ref{HV2}), which, according to our Assertion,  has a compact global attractor. Therefore, in this case the modified model with extinctions  also has a compact global attractor. 

The model with extinctions  exhibits a highly stochastic behavior. The final population state 
depends dramatically on initial data (\ref{Idata}) .\cite{Vak16} For some initial 
abundances all species coexist, whereas for other initial data only a 
few species can survive over long timescales. Usually, the environmental  influence 
diminishes the number of surviving species. Nonetheless sometimes
the environmental chaos can stabilize the ecosystem, increasing the number of 
coexisting species.  Systems with large numbers of species
are stabler than ones with few species. This multistability, which is present in a system with fixed parameters, means that in a system with slowly evolving parameters we can observe
jumps between equilibria.

\section{A more detailed look at extinctions in our model}

We follow Kozlov et al. (2017)\cite{Vak16} but will consider the problem of extinction in more detail. 
 Let us consider the case of a single resource $M=1, v_1=v$ for large
 $D$. Let 
 $\phi_i=a_i \phi(v)$, where $\phi(v)=\frac{v} {K+v}$. Then, according to our Assertion, all trajectories are convergent to equilibria.  Let $N$ be the number of coexisting species for such equilibria and $v_{eq}$ is the equilibrium amount of the resource. 
Then the equilibrium abundances $\bar x_i$ are \cite{Vak16}
\begin{equation} \label{barx}
\bar x_i=  (\gamma_i^{-1} (a_i\phi(v_{eq})- r_i))_{+, \delta}
\end{equation}
where $a_{+, \delta}$ is truncated at level $\delta$ the number $a$: $a_{+, \delta}=a$ for 
$a > \delta >0$ and $a=0$ otherwise.  For $v_{eq}$ we then obtain 
\begin{equation} \label{veq}
D(S- v_{eq})= \phi(v_{eq}) \sum_{i=1}^N a_i c_i (\gamma_i^{-1} (a_i \phi(v_{eq})- r_i))_{+, \delta},
\end{equation}
where $c_i=c_{i1}>0$ are  coefficients and we assume that $c_i > c_0 >0$. 
Note that $v_{eq}$ depends on $S$ and $N$. That dependence on $S$  is monotonic: as $S$ decreases, $v_{eq}$ also decreases. 
Together with $v_{eq}$ the equilibrium abundances $\bar x_i$ decrease and for some $i$ the value $\bar x_i$ defined
by (\ref{barx})  equals zero. Then the corresponding species suffers extinction and  the species number $N$ takes
a smaller value, for example, $N-1$. That is a typical
picture for general $S$ and not too large $N$.  

To understand how extinctions occur in our conceptual model, we consider the case of the maximal biodiversity. To simplify our analysis we suppose first that all species have identical properties, i.e, all $r_i=\bar r$
and $a_i=a$,  $c_i=c$, $\gamma_i=\gamma$.  Then from (\ref{veq}) one has

\begin{equation} \label{Neq}
N= D\gamma a^{-1} c^{-1} \frac{D(S- v_{eq})}{\phi(v_{eq})(a \phi(v_{eq})- r))_{+, \delta\gamma}}.
\end{equation}

An analysis of that equation  allows us to note that in (\ref{Neq}) the numerator decreases in $v_{eq}$ and the denominator is an increasing function of  $v_{eq}$. Thus $N$ is a decreasing function of the $v_{eq}$. We conclude that there holds a
 
 {\bf Principle of resource minimum}
 {\em   The maximum of biodiversity  is achieved with a minimum of resources.  }

Consider now how extinctions can occur. 
While $a \phi(v_{eq})- r >> \delta \gamma$, a small variation $\Delta S$ in the resource $S$ leads to a
small variation in $N$, typically $N$ conserves. In fact, a decrease in $S$ can be compensated by the corresponding decrease   of the consumed resource amount $v_{eq}$. In this case we observe the extinction of a small number of species.


However, in the case of the maximal possible biodiversity $N$ that can be attained if all $\bar x_i $ are close to $\delta$, the situation dramatically changes when the
resource $v_{eq}$ is also close to the maximal value
$S$. In this situation, a decrease in a $v_{eq}$ leads to extinction of many or even all species in the model because for smaller $v_{eq}$ we have
$a \phi(v_{eq}) < r +\delta \gamma$. 

\begin{figure*}
\centering
\includegraphics[width=1.0\linewidth]{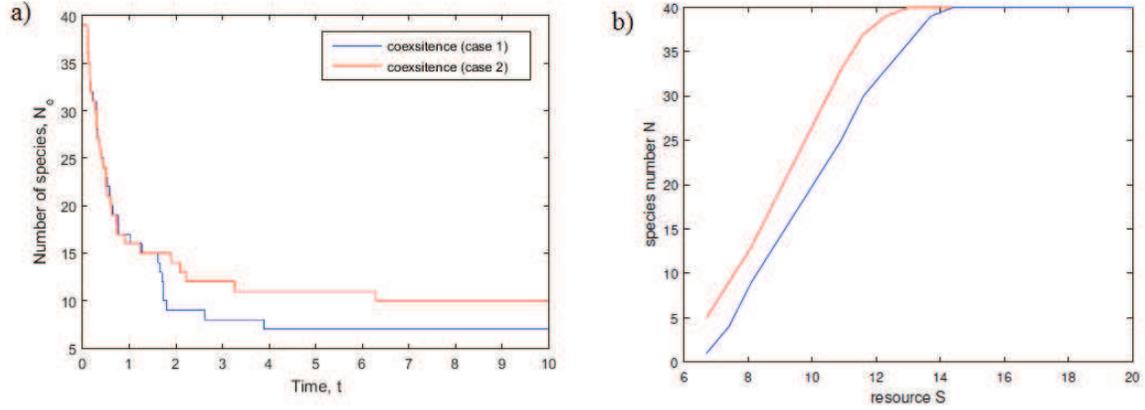}
\caption{a) This graph shows the number of coexisting species $N_e(t)$ in the two cases. In case $1$, the number of species $N_e(t)$ coexists 
 in a system depending on $t$ when the nutrient supply $S$ is  a constant, $S_0=200$ and
$S_1=0$.   
In the case $2$, the number of species $N_e(t)$ coexists in a system with time varying resource, $S_0=200, S_1=180$. 
b) This plot shows the dependence of the number 
of coexisting species (vertical axis) on $S$ (horizontal axis) for a random population composed 
as follows.  The parameter $K=4$ and $\gamma_i=\gamma=1$. The parameters
$a_i$ are chosen according to log-normal law with average $1$ and $\sigma=0.03$.
The mortality parameter $r_i=R$ is chosen so that the species abundances are close to the limit value $\delta=1$, $R= 14$. The parameters $c_i =c(1+ \tilde c_i)$ where
$c=0.1 $ and $c_i$ are uniformly distributed on $[0,1]$.
The red curve line corresponds to the non-perturbed case and 
the blue curve describes biodiversity when the resource limit $S$ is diminished on $10$ percents with respect to the non-perturbed value.} 
\label{Fig1}
\end{figure*}

This effect is weaker if the species parameters are different (i.e. the parameters $a_i, c_i, \gamma_i$ are different). We have studied this situation numerically and the results obtained are shown in Fig.\ref{Fig1}.

From this study of our model we can formulate the following assertion:

{\bf Extinction principles}
{\em
({\bf a}) If an ecosystem consisting of species that share the same resource attains its maximum possible biodiversity, then relatively small changes in the environment (such as in the climate) can lead to species extinction. ({\bf b}) If the biodiversity of an ecosystem is at its maximal possible value and 
simultaneously the species in that ecosystem consume resources close to a maximal value, then the ecosystem is fragile: it can be destroyed completely or almost completely 
as a result of species extinction under very small environmental changes. This effect is weaker for ecosystems consisting of a random mix of species that have different mortality and resource consumption parameters. 
}

\section{The population model under periodic and chaotic environmental forcing}

In this section, we consider extinctions in our model forced by chaotic and periodic environmental temperature $T$ changes. We assume that the resource supply depends on $T$ and that $T$ is a periodic function of time. We also include stochastic effects.  
For example, we can suppose that
\begin{equation} \label{supplyper} 
   S=S_0 + r \sin(\omega t )  + \epsilon \eta(t) 
\end{equation}
where $S_0,  S_1 >0$ are parameters, $\omega$ is a frequency, $\eta$ is standard white noise and   
$\epsilon$ is the noise amplitude.  This means that 
the temperature changes periodically in time. The parameter $S_0$ represents nutrient supply (the resource available to species), and the parameter $r $ describes the intensity of periodic forcing. 
 
To simulate chaotic time forcing we set 
\begin{equation} \label{supplychaos} 
   S=S_0 + r \theta(q(t)) 
\end{equation}
where  $\theta(q)$ is a smooth function of the vector argument $q$, $q=(q_1,...,q_n)$ which describes a state of the ecosystem environment (the climate for example), and the dynamics of $q$ is governed by trajectories of the noisy dynamical system, written in the Ito form:
\begin{equation} \label{dynsys} 
   dq=Q(q)dt  +  \sqrt{\epsilon}  dB(t),
\end{equation}
where $B(t)$ is the standard Brown motion and $Q$ is a smooth vector field. In the case $\epsilon=0$ we are
dealing in (\ref{dynsys}) with a system of differential equations, and we will suppose that its dynamics are well posed and has a compact attractor ${\mathcal A}_Q$. Then for small $\epsilon$ we can use
the Freidlin-Wentzell theory,\cite{FW} and the properties of the noisy dynamical system (\ref{dynsys}) depend heavily on the attractor structure. 

For example, we can set $q=(x, y, z)$ and consider
the Lorenz system, a rough model of atmospheric dynamics given by 
\begin{equation}
\label{A.12} 
\begin{split} 
dx/dt=\tau^{-1}(\alpha(y-x)    ),\\
dy/dt=\tau^{-1}(x(\rho -z) -y  ),\\
dz/dt=\tau^{-1}(xy - \beta z  ),
\end{split}
 \end{equation}
where $\alpha, \beta, \rho$ are parameters,  and $\tau>0$ is a parameter that controls the 
speed of the trajectories.
For $\epsilon=0$ that system shows a chaotic behaviour for $\alpha=10, \beta=8/3$ and $\rho=28$.
We construct $\theta$ as follows.  The third component $z$ in (\ref{A.12}) describes the time evolution of temperature.
We set $\theta(t)=(z(t)- \bar z)/\mu_z$, where $\mu_z=\max (|z(t)|)$ on a large interval $[0, T]$ and 
$\bar z$ is the average of $T^{-1} \int_0^T z(t)dt$ on this interval.

\begin{figure*}
\centering
\includegraphics[width=1.0\linewidth]{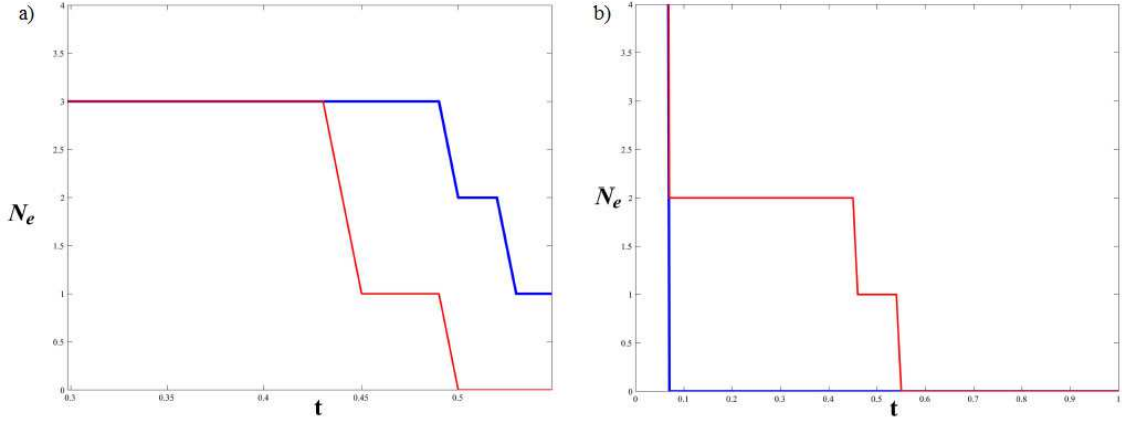}
\caption{The plots show the dependence of the number 
of coexisting species (vertical axis) on time (horizontal axis).
The blue line corresponds to the case when all parameters 
do not depend on time
and periodic environmental forcing does not influence the system; the red line 
describes time evolution under periodic environmental forcing influence.  
a) The plot shows environmental forcing dependence diminishes biodiversity. b) The plot shows environmental forcing dependence
increases biodiversity.}  
\label{Fig2}
\end{figure*}

The time extended model reduces to the time independent model with constant $S$ in the two opposite cases:  ({\bf A}) $\omega >> 1$ and ({\bf B}) $\omega << 1$. 
Assume   $S=S(t)$ is  defined by (\ref{supplyper}). 
In case ({\bf A}), we can apply the averaging principle to   (\ref{HX2}), (\ref{HV2}) and replace $S(t)$ by $S_0$ in (\ref{HV2}).  This averaging also works 
for $S(t)$ defined by  (\ref{supplychaos}). The number $N_e(t)$ of coexisting species tends to a constant for large $t$. This asymptotic approach is confirmed by numerical
results in a large diapason of parameter values.  

In case ({\bf B})  we  introduce a slow time $\bar t=\omega t$ and use a quasistationary approximation.
Then we obtain that the equilibria $\bar x(\bar t), \bar v(\bar t)$ are functions of  slow time.  The number $N_e$ of coexisting species  is also a function of $\bar t$. Note that $N_e$ is a measure of biodiversity in our model. 

Before we present some numerical results, we will show (using methods of the Freidlin-Wentzell theory \cite{FW}) the existence of three sharply different extinction scenarios in our model, which can be generated by random and non-random climate variation induced by \eqref{dynsys}.

Our primary goal here is to find the probabilities of extinctions in our model. For simplicity, we consider the case of a single resource and denote by ${\Delta S}_c$ the critical value of supply change that leads to an extinction, and let ${\mathcal O}({\Delta S}_c)$ be the set of values $q$ corresponding
to that value:
$${\mathcal O}({\Delta S}_c)= \{q:   \theta(q)={\Delta S}_c) \}.
$$
Following \cite{FW} we define distance between $d(q, q')$ between two points
$q$ and $q'$ by
$$
d_{FW}=\inf_{p(t), p(0)=q, p(T)=q'} \int_0^T   (dq/dt - Q(q))^2 dt.
$$
The distance between the two sets $A$ and $B$ is defined as
$dist(A, B)= \inf_{q \in A, q' \in B} d_{FW}(q, q')$. We note in particular that if 
$q, q'$ lies in the same connected component of the attractor then $d_{FW}(q, q')=0$.
We assume first that the attractor consists of a single connected component. 
Then the probability $P_{c, \epsilon}$ to attain the critical value starting from a point on the attractor satisfies the estimate 
\begin{equation} \label{probest}
\lim_{\epsilon \to 0} \epsilon^{-1} \log P_{c, \epsilon} =   inf_{q \in {\mathcal A}_Q, q' \in {\mathcal O}({\Delta S}_c)} d_{FW} (q, q').
\end{equation}

Using that relation and known results \cite{FW} we obtain that 
there are three possible extinction probability scenarios as a function of $\epsilon$.

{\bf I} If the intersection $I={\mathcal A}_Q \cap {\mathcal O}({\Delta S}_c) $ is not empty for all 
 ${\Delta S}_c$ then 
the probability $P_{c,\epsilon}$ is not exponentially small, i.e., 
$\lim_{epsilon \to 0} \epsilon^{-1} \log P_{c, \epsilon}=0$. It is a catastrophic scenario when the extinction of all species is quite probable.

{\bf II}
The intersection $I={\mathcal A}_Q \cap {\mathcal O}({\Delta S}_c) $ is empty for all 
 ${\Delta S}_c$.  Then 
the probability $P_{c,\epsilon}$ are  exponentially small both for large and small extinctions.

{\bf III}
The intersection $I={\mathcal A}_Q \cap {\mathcal O}({\Delta S}_c)$ is not empty  for  some
${\Delta S}_c$ but  it is  empty for larger ${\Delta S}_c$. Then it is possible that
the probability $P_{c,\epsilon}$ is not small for extinctions involving relatively few species but that probability is exponentially small for extinctions involving relatively many species. In this case, there is a sharp transition in the probabilities of small losses of biodiversity and great losses of biodiversity.

If the attractor consists of $n_A >1$ connected components ${\mathcal A}_Q^{(i)}$ we observe that there are possible additional effects that may be caused by climate bifurcations (tipping points). 
Then the climate bifurcation can correspond to a transition from a connected component to another one that may lead for example to a transition from scenario I to scenario II (or III), and vice versa.

The numerical results for periodical and chaotic cases are as follows. 
For large  values of $S_0$ and $\omega \in (3, 8)$, when the 
period of time oscillations is much less,  system 
 (\ref{HX2}), (\ref{HV2}) with $M=1$ shows formidable stability even 
for $r$ close to $0$.  The periodic and chaotic 
oscillations always decrease biodiversity, but the effect 
on coexisting species is small: the numbers $N_e(T)$ remain 
close or they coincide. To obtain diminishing biodiversity, 
it is necessary to take $r =0.05S_0$ which
corresponds to the case of very strong oscillation.
In rare situations, the counterintuitive effect of biodiversity  
increasing under oscillations is possible. It may happen 
when the averaged resource $S_0$ is not large.
Note that this effect can be explained.  In fact, the 
time oscillations and increasing supply can conserve some 
species that were close to extinction.

Typical situations showing the dynamics of the number of coexisting 
species and how the environmental forcing changes that number, 
are illustrated in Fig.~\ref{Fig2}.
Here we assume that the $i$-th species survives 
while $x_i(t) > X_{ext}$, where $X_{ext}$ 
is a small parameter.  The initial species
abundances and all parameters are defined by  
log-normal distributions,  $\phi_i(t)$ are 
distributed by the standard normal law $N(0,1)$. Initially
the number of species is $N=100$.  
We see that initially the number of coexistence species declines rapidly. This effect has a clear interpretation: the resources can only support some bounded number of species. Further, we observe a slow extinction process, which progresses differently according to whether
there is or is not an environmental influence in the model evolution (see Fig.~\ref{Fig2}).




These plots and other numerical results can be interpreted as follows.

1) If the population is stable, i.e. all species survive, 
then periodic temporal dynamics of the environment increase the species abundances
and the total biomass. If the environment  evolves with chaotic or random components, this effect diminishes.

2) If the population experiences harsher conditions, 
environment oscillations can lead to extinction
of all species. However, if the species survive, 
the environmental oscillations can increase
biomass. 


\section{Concluding remarks}
In this paper, a conceptual resource  model for biodiversity is proposed and studied. Our conceptual model describes a simple and easily understandable mechanism for resource competition, and generalizes the well-known Huisman and Weissing model,~\cite{HuWe99} taking into account species self-regulation, extinctions, and time dependence of resources. Our numerical results show that when the averaged resource supply level is large enough, fast time oscillations in resource supply do not effect essential biodiversity (the number of coexisting species). This result is valid both for chaotic and periodic oscillations. The effect of oscillations becomes observable when the averaged resource value is sufficiently small. Then, typically, the oscillations (both chaotic and periodic) diminish biodiversity, although in some cases oscillations with a noise component can increase biodiversity.

In our model the largest extinctions occur when resource consumption reaches a maximal possible value, but there is a smooth continuum from extinctions of relatively small magnitude (the loss of a few species) to extinctions of relatively large magnitude (the loss of a great many species). Thus, we are not able to identify mass extinctions (sensu Sepkoski (1986)\cite{Sep86}) as a quantitatively different regime (e.g. Jablonski (2005)\cite{Jab05}). This is likely to be because our conceptual model currently does not include trophic levels such as primary producers, herbivores and predators, or evolutionary processes such as speciation (cf. Sole et al. (2002)\cite{Sol02}). Similarly, our analyses have focused on the conditions that lead to extinction. Representation of ecological structure and evolutionary processes such as these in future extensions of our model will allow us to investigate the dynamics of recoveries from extinction, and this will permit investigations of how ecosystems rebuild and new ecologies emerge from the aftermath of extinction events.  
 
Nevertheless, our conceptual model provides support, on theoretical grounds, for the importance of non-linear processes during the various extinction events that have punctuated Earth history. For example, the rapid loss of plant biodiversity during an extinction event in the Late Triassic period (200 million years ago) has been attributed partly to a threshold response of plants to relatively minor increases in the concentration of carbon dioxide in Earth's atmosphere at this time \cite{El05}. Additionally, when ecosystems reach maximal biodiversity in our model, the risk of large extinction events strongly increases, even under small environment changes, and random, chaotic or periodic environment oscillations can also dramatically affect biodiversity. Thus, suggestions that the global diversity of life on Earth is capped somehow (see discussion in Benton and Emerson (2007)\cite{Ben07}) are not incompatible with the results of our conceptual modeling. 	

\section*{\small Acknowledgement}
The authors are grateful for financial support from the Government of the Russian Federation through the Mega-grant No.074-U01. We also acknowledge support from the the Russian Foundation for Basic Research (RFBR) under the Grants No.16-34-00733 mol\_a\ and No.16-31-60070 mol\_a\_dk. In addition, we gratefully acknowledge support from the Division of Physics at the U.S. National Science Foundation (NSF) through Grant No. PHY-1066293. Finally, we thank the Statistical and Applied Mathematical Sciences Institute (SAMSI) and the Mathematical Biosciences Institute (MBI) for their support of this work.

\bibliographystyle{plain}

\end{document}